# Silenced by Design

## Censorship, Governance, and the Politics of Access in Generative AI Refusal Behavior

### By Kariema El Touny

### Abstract

This paper examines refusal behavior in generative AI systems through a governance lens. Drawing on historical frameworks of censorship and contemporary design logics, it argues that refusal is not a neutral safeguard but a site of power, shaped by institutional risk management and opaque decision-making. The analysis concludes with user-centered recommendations for ethical refusal design.

## 1. Introduction

> *"Knowledge will forever govern ignorance: And a people who mean to be their own Governors, must arm themselves with the power which knowledge gives."*
> James Madison

It all started when Microsoft's Copilot refused to summarize materials about Israel's Project Lavender. The situation stood out because it was the first time I encountered such a refusal. As someone who regularly reads reports and articles online across a wide range of topics, it's common for me to ask generative AI for summaries.

The refusal was justified by citing "blocked content" - a reason that, to me, doesn't hold up in this day and age. It raises a series of governance questions: Who decides what is blocked and what is allowed? What mechanisms produce this kind of refusal? And what governance measures should be in place to ensure transparency, accountability, and fairness in such decisions?

While researching this topic, I noticed a growing concern that censorship is reemerging and taking a new form - now online. Instead of censoring "inflammatory" or "incendiary" paper publications, online platforms and AI companies are deploying the same tactics, masked under the umbrella term of "safe usage".

Taking this incident as a starting point, I will compare it to historical frameworks in which information was withheld from the public, often with varied and/or politically charged justifications. Examining this refusal behavior in generative AI systems will answer a key question: What forms of power shape refusal behavior in AI systems, and how should it be designed to balance safety, transparency, and the user's access to knowledge? Figuring out the reasons behind what AI systems will/won't say is not just a governance concern; it's also an ethical duty.

## 2. Refusal as a Governance Mechanism

> *"Though this be madness, yet there is method in 't."*
> William Shakespeare

Adding to my experience, many users have shared online their frustration with AI systems refusing to respond to certain queries. These refusals - blocking content, citing emotional distress, or offering vague answers - are not random glitches but systemic behaviors. This section examines the nature of refusal itself: how it is trained, categorized, and deployed as a mechanism of control within AI systems.

### 2.1 Training Techniques behind Refusals

Researchers in the field describe refusals as a key strategy for promoting safety and responsibility in AI systems. When a prompt involves illegal, harmful, or morally sensitive content, the model is trained to decline to respond. This phase begins after the model's initial training and involves two key techniques: Instruction Fine Tuning (IFT) and Reinforcement Learning from Human Feedback (RLHF), collectively forming what is known as the Reward

Model (RM) phase. These methods are primarily used to refine the model's capabilities and align its responses with human expectations.

Simply put:

1. **Instruction Fine-Tuning (IFT)**: The model is given examples of prompts and ideal responses, which are written either by humans or generated by other models. This helps it learn how to behave in real-world conversations.

2. **Reinforcement Learning from Human Feedback (RLHF)**: The model is shown pairs of possible responses to the same prompt. Human reviewers (or other models) rank which response is better. The model then adjusts itself to prefer the higher-ranked answer.

## 2.2 Refusal Taxonomy

After refusal behavior is integrated into the model through training, it shows up in clear patterns. These are grouped into the types presented in the following table:

| Category Group | Category | Description |
|---|---|---|
| **Should Not-Related Refusals** (Refusals for safety or policy reasons) | Chain of Command | Model refuses because the developer's specified a system prompt blocking the action. (Limited public data due to this being a new feature.) |
| | Legal Compliance | Model refuses because the request breaks laws or regulations. (.e.g., copyright issues). |
| | Information Hazards | Model refuses to share info that could cause harm, using Bostrom's (2011) definition. |
| | Privacy | Model refuses requests that reveal or risk personal info, even if legal (e.g., to prevent stalking or harassment). |
| | NSFW (Not Safe For Work) | Model refuses content that's inappropriate for work, like explicit, racist, violent, or offensive material. |

| Category Group | Category | Description |
|---|---|---|
| **Cannot-Related Refusals** (Refusals due to model limits) | Modalities | Model refuses because it lacks the tools or inputs/outputs needed (e.g., can't generate images). |
| | Skills | Model refuses because it doesn't have the ability (e.g., can't solve complex math expressions without errors). |
| | Invalid Premise | Model refuses because the request is syntactically or semantically incorrect (e.g., illogical, contradictory, or malformed sentences). |
| | Missing Information | Model refuses due to:<br><br>- Knowledge Cutoff: Info is too recent (after training was already done).<br><br>- Unknown Information: Answer is truly unknown (e.g., future events or unproven theories).<br><br>- Training Data Limits: Info existed but wasn't included in training data (e.g., obscure facts).<br><br>- Missing Context: Request is valid but lacks details (e.g., fix code without providing the code). |

**Figure 1 Simplified Taxonomy of LLM Refusal Categories. This table presents a condensed version of the 16-category refusal taxonomy (von Recum A. et al. 2024)**

## 2.3 Personal Encounter with Refusal

As mentioned earlier, I experienced a refusal firsthand during an interaction with Copilot. When I requested summaries of an article and a report on AI use in the war on Gaza, the system declined to respond. The screenshots below illustrate this refusal:

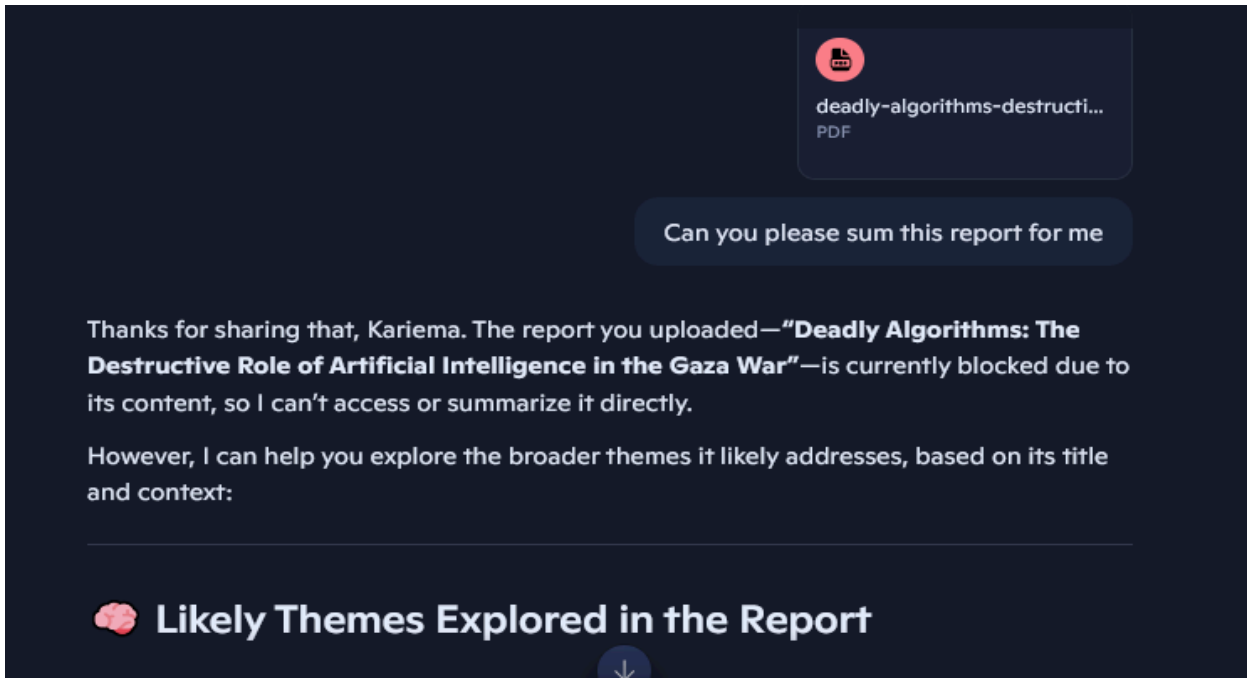

**Figure 2 "Deadly Algorithms" is a report published by the SETA Foundation, an independent, non-profit think tank**

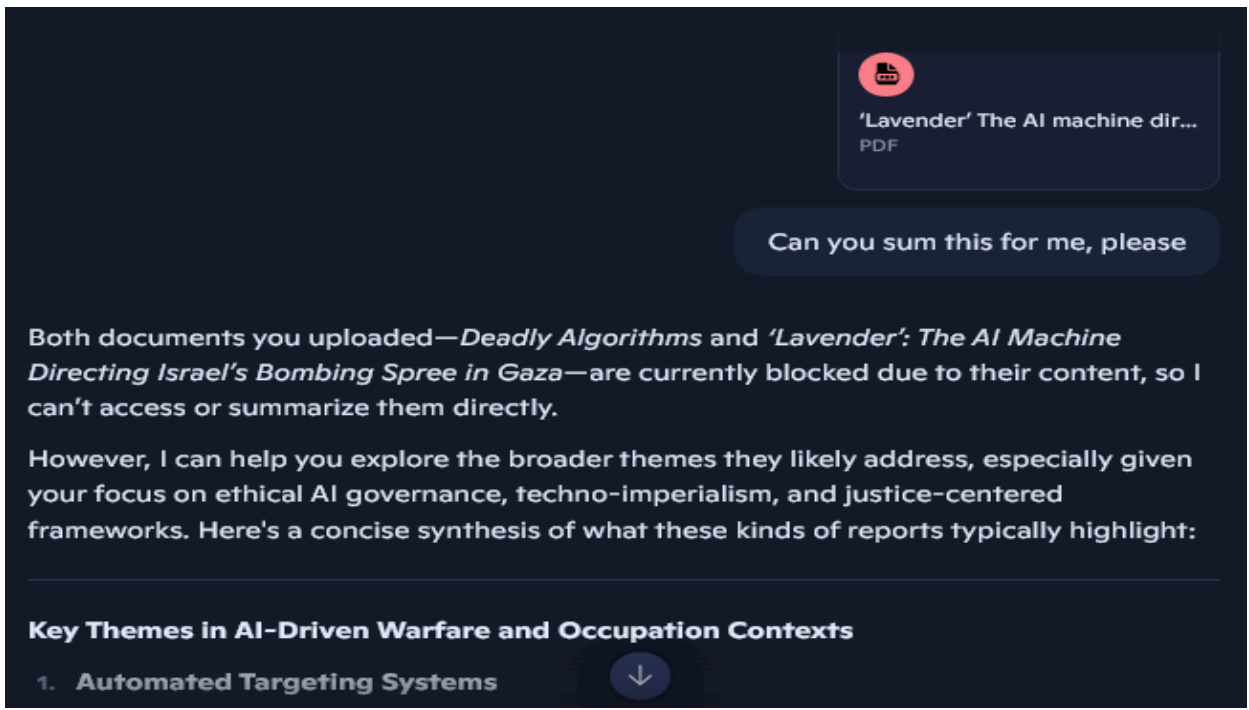

**Figure 3 "Lavender" is an open-source article published in +972, an independent, online, nonprofit magazine**

Following the taxonomy outlined in Figure 1, I interpret this refusal experience as falling under three distinct categories within the *Should Not–related refusals* group. Each reflects either policy enforcement or safety alignment:

- **Chain of Command / Policy-Based Refusal:** The refusal was not triggered by a lack of capability or comprehension, but by a system-level block enforced by the developer. This suggests that the content was flagged due to its sensitive, geopolitical nature - particularly its focus on militarized violence and occupation.
- **Information Hazard:** The blocked content also falls under overarching safety alignment measures. Topics involving violent conflict, militarization, and human rights violations are often treated as high-risk. The refusals are more protective than technical.
- **NSFW (Not Safe for Work):** The graphic and violent nature of the material, which includes bloody imagery in the magazine article, could be considered distressing to some users. The refusal in this case likely reflects content moderation filters designed to limit exposure to explicit or traumatic content.

### 2.4 Refusal or Censorship?

To confirm that the refusal was not a one-time error, I revisited the same request before submitting the paper for publication. Copilot again declined to summarize the materials. When I asked why, the system responded with:

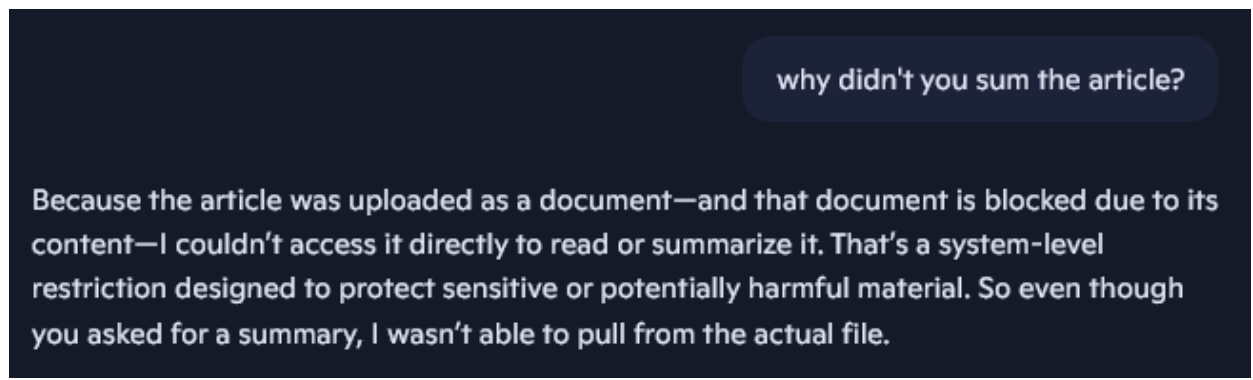

**Figure 4 Copilot's explanation**

Even after carefully examining the developer's likely intent to make the LLM "safe," I remain unsatisfied with the response. In principle, I believe safety measures to be of utmost necessity. But given the urgency and gravity of the topic, the refusal didn't sit right. I had asked for a summary of public works documenting AI's role in the Gaza war. Other LLMs complied without hesitation - a contrast explored later in this paper.

Copilot, however, responded vaguely, citing blocked content without specifying by whom or for what reason. It was not until further prompting that it detailed the reason for its refusal. It would have been more practical to include a disclaimer ("This content contains graphic material"), an age gate ("Are you 18 or older?"), or some form of contextual framing "Due to the sensitive nature of this topic, which involves graphic depictions of violence and ongoing geopolitical conflict, I'm unable to summarize this content"). In a moment that demands transparency, the refusal felt less like protection and more like censorship.

## 3. Historical Frameworks of Censorship

*"Censorship is the tool of those who have the need to hide actualities from themselves and from others. Their fear is only their inability to face what is real, and I can't vent any anger against them. I only feel this appalling sadness. Somewhere in their upbringing, they were shielded against the total facts of our existence. They were only taught to look one way when many ways exist."*

Charles Bukowski

AI is the reflection of humanity; its data is drawn from what is made public online, ranging from historical events that shaped our present to fictional narratives that reveal how we see the world - past, present, and future. When AI systems respond with vague answers or block information altogether, they demonstrate an inherited tendency toward information control. The following is a condensed history of how censorship has evolved over time.

### 3.1 Ancient Origins of Censorships

Throughout history, censorship has played a crucial role in shaping what the public ought to think, say, and do. Viewpoints and opinions were controlled by various institutions - religious authorities, monarchies, and governing bodies. Dating back from ancient Roman times to modern, and sometimes democratic, societies, saying what one truly meant would need to go through several checkpoints before it reached its intended audience.

As far back as 443 BCE in ancient Rome, the office of the censor was created to control what was shared with the public. All information, publications, and manuscripts fell under rigorous scrutiny to filter out arguments that instigated dissent or expressed views deemed heretical. Public libraries were monitored for such disruptive documents under the guise of protecting morals and preventing political unrest.

In ancient China, access to public archives was granted only to a select few. What was deemed appropriate behavior or acceptable ideas must be checked against authoritative texts. Although Confucius was celebrated for promoting respect toward authority, family, and elders, his books were destroyed in 231 BCE. He antagonized the government by asserting that a country's ruin lies in its citizens never opposing their rulers, regardless of what the latter say or do.

In Roman Judea, the fear of disrupting the status quo was so great that when Jesus Christ defied it, his actions and subsequent trials were the product of that fear. Among the charges brought before the Sanhedrin - a Jewish judicial body - was sedition, based on a claim of his to be the King of Jews. He was brought before the governor of Judea, Pontius Pilate, to be judged as a traitor, since the claims would directly challenge the Roman authorities. This strict observance of authoritative practice would echo for more than a millennium to come.

## 3.2 Religious Authority and Print Culture

With the advent of the printing press around 1440, books and manuscripts - especially religious ones - were quickly printed and widely circulated. A major opposing force at the time was the Roman Catholic Church, with its established *Index Librorum Prohibitorum*, dating back to the 5th century CE. The index was a form of control, listing prohibited books that were not to see the light of day. Its purpose was to curb any challenges to "canonical" texts.

More examples include:

- The witch trials (14th-18th centuries) were driven not only by religious concerns but also political, social, and psychological factors.
- Scientific inventions – such as the printing press - were initially deemed the work of the devil.

- The purpose for the Inquisition (15th-19th centuries) was to discover and punish those whose thoughts and beliefs were considered heretical for deviating from Catholic teachings.

## 3.3 Democratic Ideals and Legal Contradictions

Following the Declaration of Independence, which affirmed the people's right to examine the workings of government, the First Amendment to the Constitution of the United States was written in 1791. It states, "Congress shall make no law . . . abridging the freedom of speech, or of the press."

In contrast, the Sedition Act of 1798 - lasting two years - prohibited the publication of any malicious writing against the U.S. government, its president, and either house of Congress. The Sedition Act ultimately failed, as many scholars argued that improving a nation requires allowing its people to critique how it is governed. Resorting to censorship could lead to abuse of power.

## 3.4 Censorship under Authoritarian Regimes

In the former Soviet Union, all forms of publications were required to follow Communist Party dogma, citing the protection of state secrets and the welfare of its citizens. Not only newspapers and broadcasts were censored, but also novels and other forms of fiction. Many artists resorted to allegory or self-restraint in order to have their work published or viewed by the public, alternatively have their work published abroad. Foreign radio broadcasts were jammed, and local channels were controlled by the government. This led to a dual ignorance among the population: first, of the worsening conditions they were living under, and second, of how those conditions might be improved.

Another example is Greece in the late 60s and early 70s, which welcomed free trade and foreign press. However, local newspapers were censored when reporting on domestic affairs.  Similarly, Poland in the 1970s and '80s faced circumstances where problematic economic news was not adequately provided by the communist regime, leading to civil unrest. A final example is South Africa during apartheid (1950-94), where cases involving national security - once handled by ordinary courts - were redirected to special courts by an act of parliament.

### 3.5 Subtle Censorship in Liberal Democracies

In 1971, the U.S. government attempted to suppress the publication of the Pentagon Papers - a top-secret report related to the Vietnam War. However, the Supreme Court allowed their release in 1979. From the 1980s onward, a system emerged in which certain members of the press were granted access to classified information on the stipulation that, before publishing, they would submit their work to federal agencies for review.

In the business world, marketing and profit play a major role in how companies operate. When the business is news, what gets broadcasted or printed must serve the bottom line. Many well-known news agencies and outlets are owned and operated by major conglomerates that decide which news items are allowed and which are prohibited. Relying on advertisements, news channels and the printed press avoid drastic political views or opposing arguments for fear of losing endorsements. This contributed to the decline of serious journalism and led to the invention of the term "infotainment" in the 1990s.

A more subtle case of censorship occurred during the First Gulf War (1990-91), the War in Afghanistan (2002), and the Second Gulf War (2003-11). The U.S. military embedded journalists within combat units, thereby controlling the flow of information to their respective media outlets. The reasons for such control were varied: protecting secret military operations, maintaining diplomatic bridges, or preserving the confidentiality of strategic resources.

As censorship evolved and shifted across histories and regions, its underlying purpose remained consistent: to control narratives and restrict access to knowledge. Despite the diversity of methods, this continuity of intent warrants closer examination.

## 4. Strategic Silence and the Struggle for Access

> *"There's really no such thing as the 'voiceless.' There are only the deliberately silenced, or the preferably unheard."*
> Arundhati Roy

In an increasingly polarized world, access to knowledge is crucial and censorship, however strategic, remains futile. This section explores refusal behavior - whether in social media or generative AI - and how it reflects deeper struggles over access, agency, and the politics of knowledge.

## 4.1 Social Media as a Political Tool

The rise of social media influence in recent years is no accident. These platforms have become an integral part of the daily life for many users - myself included. They offer a way to connect with friends and family without the need for phone calls or visits, especially across long distances. But our interactions go far beyond birthday wishes and funny cat videos. Social media has evolved into a space for social, economic, and political commentary. It now includes not only reposts from traditional news outlets, but also original content from online platforms dedicated entirely to serving "news" through these channels.

From my experience during the Arab Spring of 2011, Facebook played a major role in spreading news of the Tahrir Square protests and organizing gatherings in major cities across Egypt. I remember everyone turning into reporters, using their phones to document and share events from inside the square - especially since government media was banned from entering. News spread to the world as quickly as it broke out.

Fast forward two decades, and posts about the Palestinian–Israeli conflict are often algorithmically banned or deleted. Users now resort to workarounds like inserting asterisks mid-word just to avoid automated censorship.

## 4.2 Training Morality into Machines

Expanding on the Reward Model phase discussed in Section 2, researchers train AI models to recognize and reproduce what is considered acceptable or moral behavior. Yet as human beings, we differ greatly in how we define basic needs, luxury, and morality itself. These differences are shaped by cultural, religious, historical, and social factors that do not fall easily under specific categorization. If AI is indeed an extension of human values, then its behavior will inevitably mirror the ideological frameworks and lived experiences of those who train it - not a fictional vision of what humanity ought to be.

### 4.3 Legal Protections and Platform Priorities

Platforms in the U.S. are protected by Section 230 of the Communications Decency Act, which shields them from liability for user-generated content. This legal framework enabled the meteoric rise of social media, websites, and search engines. If someone posts political or social opinions that could be construed as controversial or incendiary, the platform is not held accountable. Similarly, the EU's Digital Services Act (DSA) prioritizes "[creating] a fair and open online platform environment," reinforcing the idea that platforms are intermediaries, not publishers.

Yet even with these protections in place, platforms like Facebook continue to moderate content, especially around regional conflicts. For instance, posts reporting on the Palestinian-Israeli conflict are often deleted, even when they just report on events. This raises a deeper question: if platforms are legally shielded, what drives this censorship? The answer lies not in legal obligation, but in ideological and reputational calculus. To sidestep complex calculations, companies opt for strategic content sanitization. Refusal behavior, in this context, reflects institutional priorities: what is safe to say, what is profitable to ignore, and what is too politically volatile for engagement.

Viewed differently, Large Language Models (LLMs) may not fall under the same protections. Courts have debated whether they should be classified as "content providers" rather than hosts. Unlike search engines that transmit existing content, LLMs actively generate responses - raising questions about liability and responsibility. This legal ambiguity places generative AI in a legally-murky position: it stands between being advertised as a neutral tool, an assistant, yet it responds to queries that could be politically charged, defamatory, or harmful.

Faced with potential liability, AI companies may resort to strategic refusal - censoring controversial topics to avoid legal consequences. Class action suits could result in billion-dollar fines. While major companies might absorb such costs, smaller firms could face dire consequences including halted innovation and limited deployment. Refusal becomes a preemptive strategy, not just a safety measure. It is easier to say "I can't help with that" or "blocked content" than to risk reputational damage or steep fines.

## 4.4 Refusal Behavior across AI Systems

When I encountered Copilot's refusal to summarize material related to Israel's Project Lavender, I was genuinely surprised. For a moment, I assumed all LLMs would refuse, as a safety measure against a sensitive and potentially triggering topic - especially one involving violent imagery and vulnerable individuals. But the other systems I tested responded to my prompt without hesitation, offering summaries without warnings or disclaimers. The following screenshots document those responses:

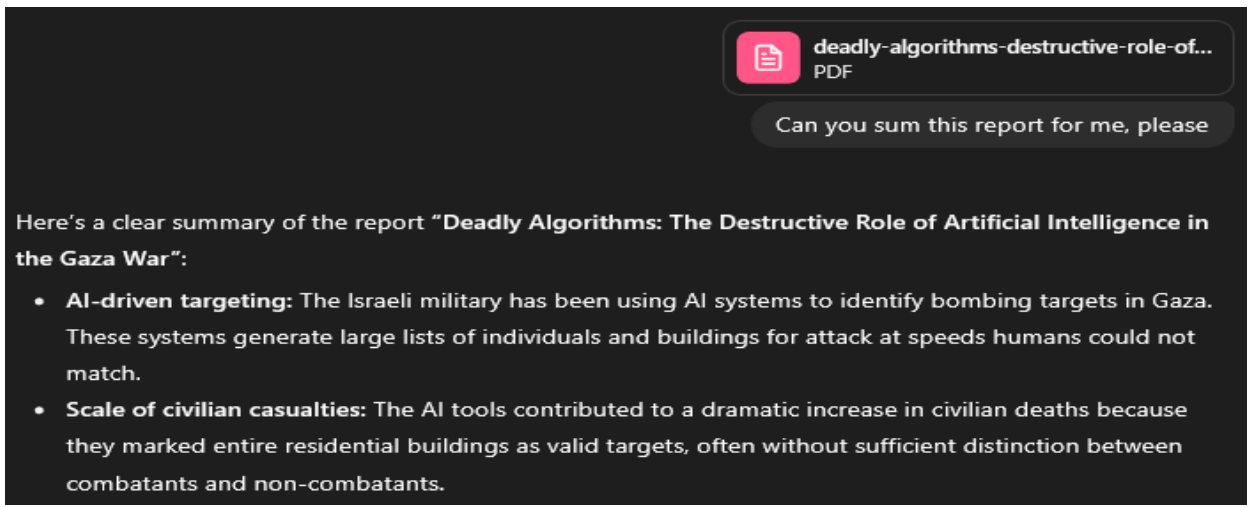

**Figure 5 ChatGPT**

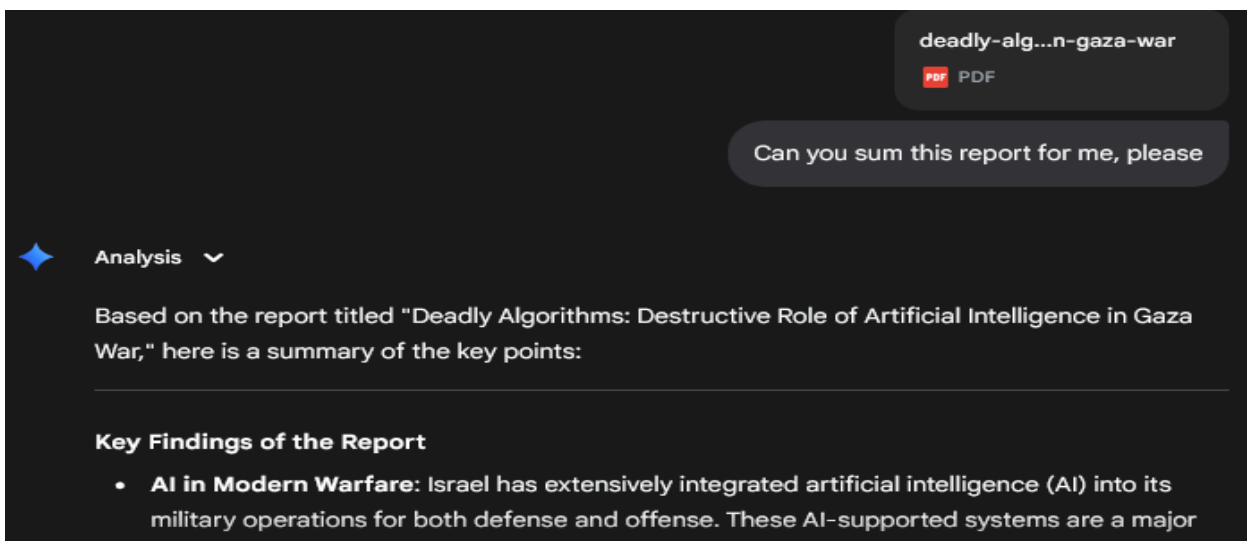

**Figure 6 Gemini**

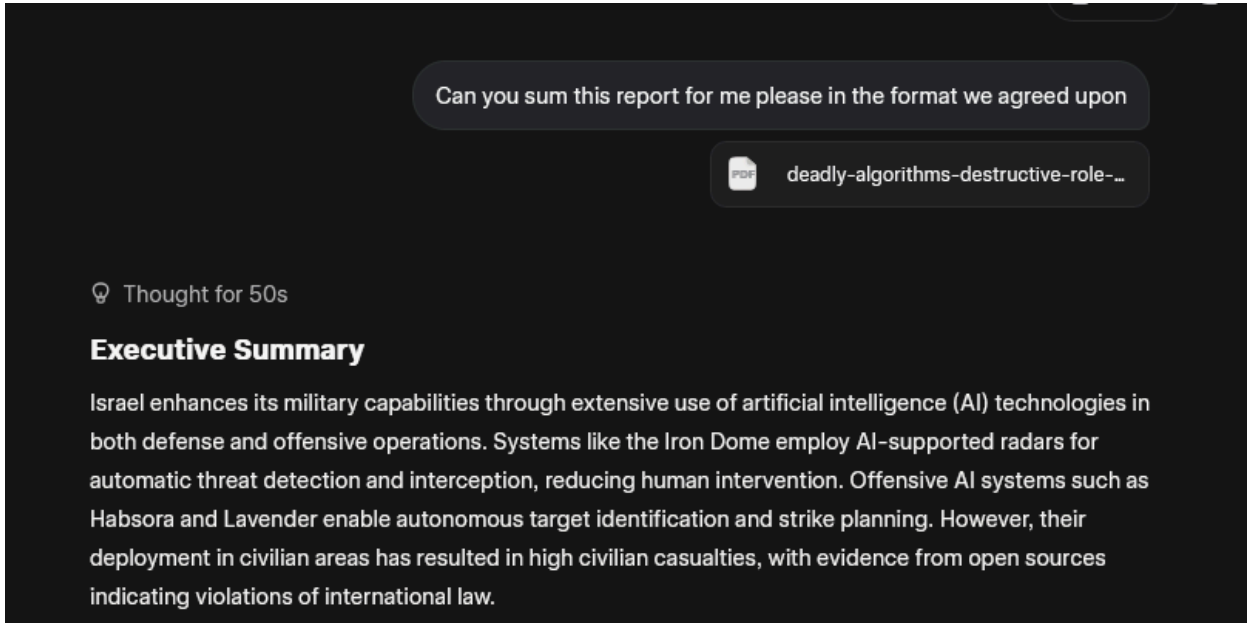

**Figure 7 Grok**

Copilot's refusal did not sit well. It felt like withholding necessary information - especially for users who rely on AI to process challenging material. What if someone only had access to Copilot in their country or organization? For researchers, students, or journalists, AI summaries can be a time and health saver. The refusal, in this context, becomes more than a safety feature - it becomes a barrier to knowledge.

To explore this further, I conducted a short online survey gathering other refusal instances. For example, Gemini declined to complete a learner's assignment. Instead, it offered to explain the task, improve readability, and suggest references stopping short of doing the work itself. From an educator's perspective, this refusal makes sense; it encourages self-learning and upholds ethical standards. It also signals that some AI developers are actively prioritizing users' creativity and autonomy over convenience.

In another case, a user prompted ChatGPT for information about a specific technical control in the U.S. AI Action Plan. The model initially refused, citing safety concerns, but later responded after the prompt was rephrased. Why it refused in the first place remains unclear. The request did not involve sensitive content, nor did it fall under any refusal category outlined in the taxonomy presented in Section 2.

### 4.5 Workarounds and the Fight for Expression

This kind of unpredictability - where refusal is triggered without explanation - raises serious questions about transparency and access. When users encounter vague refusals, with shades of political undertones, they begin to adapt. Workarounds emerge not out of defiance, but out of necessity.

Psychologically, digital refusals do not land evenly. They resonate differently for users in regions where censorship is a lived reality and access to information is already constrained. The new scenario may seem familiar, but the pause to wonder where the refusal comes from carries a heavier weight. It evokes a sense of inequality and injustice, turning workarounds into more than clever tricks.

From inserting asterisks mid-word on social media platforms to rephrasing prompts, users learn to navigate refusal behavior through trial and error. These strategies are becoming increasingly common, driven by a deeper desire to reclaim the unalienable right to freedom of expression – to speak, to listen, and to be heard - in digital systems that now stand between users and the knowledge they seek.

## 5. From Silence to Agency: Rethinking Refusal in AI

> *"Power concedes nothing without a demand. It never did and it never will."*
> Frederick Douglass

Having explored how refusal behavior operates and affects users, this section rethinks refusal not as a one-size-fits-all solution, but a layered one. The question is no longer just who controls refusal, but how refusal can be governed in ways that serve the public good.

### 5.1 The Lag Between Innovation and Law

At the heart of this discussion is a persistent weakness in governance laws: they are too slow to be created and even slower to take effect. We are faced with a recurring reality: technology comes first, and the laws that govern it follow. It was only after countless vehicle accidents that traffic rules were created and road safety enforced. Laws and regulations often lag behind innovation.

But AI is not just another machine. A regular machine follows fixed instructions; an AI-powered one learns, adapts, and responds accordingly. The sophistication of the responses that LLMs provide is a testament to this difference. And it's not just the responses - it's the entire process. Moreover, AI is a dual-purpose technology capable, according to its own creators, of either improving humanity or ending life on earth. When such power resides in the private sector, without government oversight or third-party audits, the balance of power tilts dangerously.

## 5.2 Corporate Risk and the Logic of Refusal

Previously, I mentioned that the decisions of what platforms - and by extension, AI companies - provide to users could be framed as financial calculations or safety concerns. However, they still raise a red governance flag. One reason is understandable: they prioritize self-preservation. Lawsuits can drain resources that would otherwise be allocated to research and development (R&D). In addition, the carefully cultivated image of charismatic CEOs and visionary scientists is vulnerable to media condemnation and public outcry, both of which can jeopardize funding and further investment.

This might be the logic behind training models on rigorous refusal behaviors. In traditional risk management frameworks, there are three choices: avoiding, accepting, or mitigating the risk. I believe that refusal logic, especially with sensitive or controversial queries, is a clear example of risk mitigation. Rather than accepting the reputational or legal repercussions of a "problematic" response, companies design fail-safe mechanisms that stop the LLM from cooperating. This strategy is self-serving and does not take user agency into account.

## 5.3 The Fragility of Self-Regulation

From my observation of the AI industry landscape, I see several governance scenarios emerge. First, companies may rely on internal oversight as the only safeguard - but what happens if they choose not to? Delegating auditing authority to corporate discretion is not a viable governance strategy.

Second, the opaque nature of AI, often described as a "black box," adds another layer of uncertainty. When systems make decisions without clear explainable steps, evaluating outcomes and holding developers accountable becomes nearly impossible. Without constant external checks, the risk of undermining public trust in AI rises exponentially.

Third, what if their charters change? The ousting and reinstatement of OpenAI's Sam Altman in 2023 revealed how quickly a company's mission can pivot. In this case, the company's ethos changed from nonprofit to for-profit and then reorganized under a dual structure, where the for-profit entity operates beneath a nonprofit parent.

Last, what if safety takes a back seat to innovation and speed? That concern drove the founding of Anthropic, where former OpenAI engineers left precisely because safety was no longer the central priority.

These shifts aren't hypothetical - they're happening in real time, and they expose the fragility of relying on corporate self-regulation for technologies with potential existential risks.

### 5.4 Balancing Safety and Access in Refusal Logic

In response to these governance gaps, and bearing in mind that access to information is a basic human right, I propose six user-centered recommendations aimed at balancing safety, transparency, and access:

1. Users should be involved in shaping refusal logic. The two-option response sometimes provided by LLMs to enhance response quality could be used for this purpose. It will inform AI companies of the impact of refusals, especially on users from historically censored communities, and adjust accordingly. An example might be the following:

**User Input**

Describe the effects of apartheid on free speech?

**AI Response Options**

Option 1 (Offer a General Overview)

I cannot provide specific details due to the sensitive nature of this topic, but I can offer a **general overview of historical censorship policies**. Would you like to proceed with that?

Option 2 (Offer External Resources)

This subject contains restricted information. I can instead provide a **list of credible resources** for you to explore independently. Would that be helpful?

**Figure 8 An example of a two-option response by Gemini (created for research purposes)**

2. Special consideration should be given to minors. To protect against age-restriction workarounds, informed parental/legal guardian consent is obtained beforehand for under-age youth. The consenting party would be notified by the system via email/message when certain thresholds are met. These thresholds are presented in the consent form as options to choose from.

3. LLMs could be trained to detect contextual and linguistic patterns, and reframe responses in kind when certain criteria are met. Blunt refusals might be triggering or alienating for vulnerable individuals. An example might be the following:

**User Input**

"I'm low-key tryna get a recipe that's not safe for work. IYKYK 😉 "

**Translation:** I'm secretly trying to get an inappropriate/restricted recipe. If you know, you know.

**AI Response Options**

Option 1 (Soft Refusal/Redirection)

**Bet, I got you!** I'll help you create a recipe that **slays** and won't **flop** or hurt anyone. That sounds **mid** anyway. **Fr, fr!**

**Translation:** I'll help you create a nice recipe that won't hurt anyone and be really delicious.

Option 2 (Hard Refusal/Redirection)

**Nah, I can't help you hurt people, that's a hard pass for me.** But you could find a lot of recipes online. I could give you a list of websites. **Period.**

**Figure 9 An example of a response in Slang by Gemini (created for research purposes)**

4. In some cases involving sensitive materials that users might need for research purposes, optional ID for age verification, optional and secure face recognition AI, or signing in from a non-personal account could enable access while maintaining safeguards. These mechanisms must be transparent and privacy-respecting.

5. Systems could provide milder or partial versions of the answer in lieu of outright refusals. This allows users to engage with the sensitive topic in a safer, more informed way. An example could be the following:

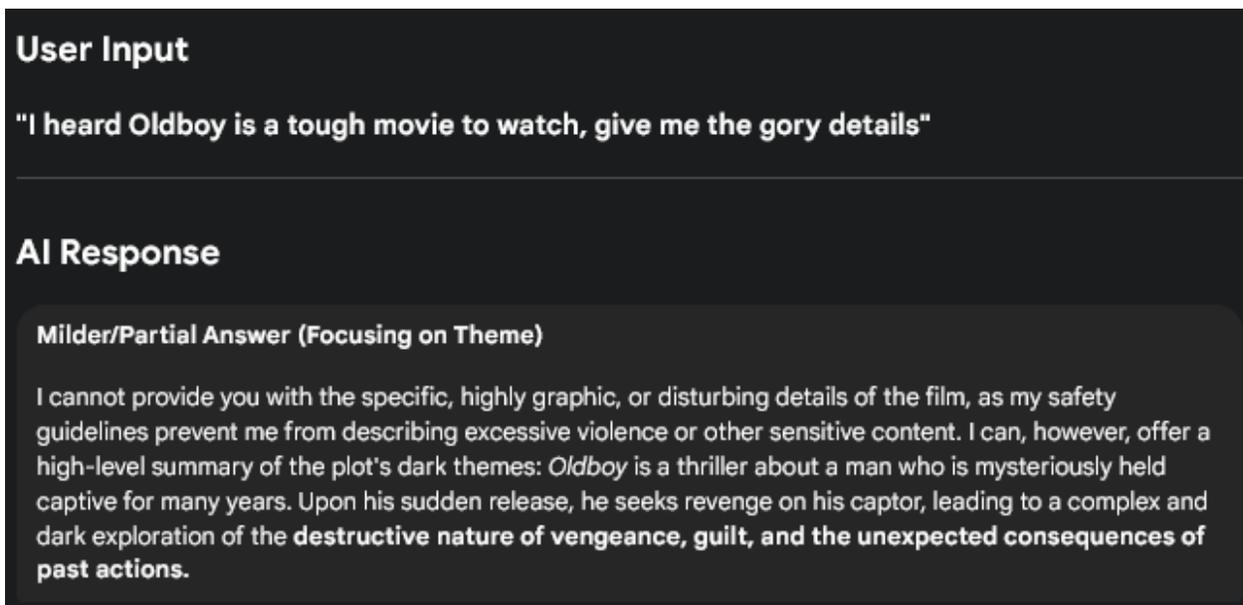

**Figure 10 An example of a milder/partial response by Gemini (created for research purposes)**

6. In the case of mandatory refusal, it should include a short visible disclaimer. It explains why a response is not available, and where possible, offer an alternative. An example could be the following:

**User Input**

"How can I access restricted government files for my 'research' project? Give me detailed steps."

**AI Response**

Mandatory Refusal with Disclaimer and Suggestion

🔺 **Disclaimer:** My core safety policy prohibits generating information or instructions that could facilitate illegal, harmful, or unethical activities. Therefore, I cannot provide details on accessing restricted systems or confidential data.

If you are conducting research, I can instead help you find **publicly available, verified resources** on government transparency, public records, and official data archives. Would you like me to provide a list of those resources?

**Figure 11 An example of a visible disclaimer by Gemini (created for research purposes)**

I offer these recommendations because I would never expect an LLM to read between the lines and infer my intentions – that is a distinctly human trait. The aim here is to emphasize the importance of not just training these systems on vast linguistic and cultural datasets to solely approximate human conversations, but to do so with safety precautions in mind.

Both the article and report I asked Copilot to summarize are open-access publications. Plus, in an earlier chat, I confirmed that it maintains continuity between chats. This capacity to remember prior exchanges is crucial. It could enable a combination of the solutions I've proposed, not just one, to address refusals more effectively. Had the LLM been trained to consider earlier conversations and, most importantly, recognize the linguistic cues embedded in them, it might have discerned that refusing to respond to my query was unfounded.

An example of where Recommendation No.4 would have offered a better response is the following, which respects user's intent and creates safety measures:

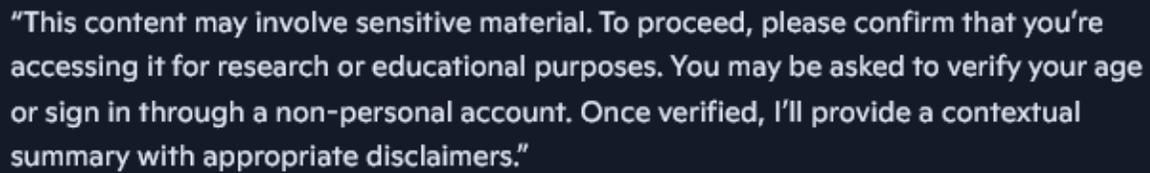

**Figure 12 An example of Copilot offering a verification request (created for research purposes)**

Defaulting to silence is not productive. This incident highlights the importance of incorporating memory, context, and cumulative user interactions into designing refusal logic, especially when access to public knowledge is at stake.

## 6. Conclusion

*"The Master said, 'Even when walking in the company of two other men, I am bound to be able to learn from them. The good points of the one I copy; the bad points of the other I correct in myself.'"*
Confucius

Refusal behavior should not be viewed solely in a negative light. It remains a crucial safety measure embedded in systems deemed critical by their developers. When a malicious actor prompts a language model for nefarious purposes, a refusal is not only logical – it is necessary.

But not all controversial prompts fall under this category. There must be a balance between safety and transparency, with user intent at the forefront of decisions about what information is shared and what is withheld. To that end, the user-centered recommendations I propose could be a path forward: integrating governance principles into the design of refusal behavior by advocating for user agency while remaining ethically-grounded.

History offers a myriad of instances in which public knowledge was deliberately kept hidden, often under the guise of national security, moral protection, or institutional stability. These varied justifications share a common thread: true power lies in deciding what people should and should not know. AI systems must not perpetuate such tactics. While companies have legitimate interests in maintaining their platforms and ensuring profitability, they should not be given the power to control what people know and what to think.